\pdfoutput=1
\RequirePackage{ifpdf}
\ifpdf
\documentclass[pdftex]{sigma}
\else
\documentclass{sigma}
\fi

\numberwithin{equation}{section}
\numberwithin{example}{section}
\numberwithin{proposition}{section}
\numberwithin{conjecture}{section}

\begin{document}

\newcommand{\arXivNumber}{1502.02948}

\allowdisplaybreaks

\renewcommand{\thefootnote}{$\star$}

\renewcommand{\PaperNumber}{046}

\FirstPageHeading

\ShortArticleName{On the Integrability of SUSY Versions of the Structural Equations}

\ArticleName{On the Integrability of Supersymmetric Versions\\ of the Structural Equations for Conformally\\ Parametrized Surfaces\footnote{This paper is a~contribution to the Special Issue on Exact Solvability and Symmetry Avatars
in honour of Luc Vinet.
The full collection is available at
\href{http://www.emis.de/journals/SIGMA/ESSA2014.html}{http://www.emis.de/journals/SIGMA/ESSA2014.html}}}

\Author{S\'ebastien BERTRAND~$^\dag$, Alfred M. GRUNDLAND~$^{\ddag\S}$ and Alexander J. HARITON~$^\S$}

\AuthorNameForHeading{S.~Bertrand, A.M.~Grundland and A.J.~Hariton}

\Address{$^\dag$~Department of Mathematics and Statistics, Universit\'e de Montr\'eal,\\
\hphantom{$^\dag$}~Montr\'eal CP 6128 (QC) H3C 3J7, Canada}
\EmailD{\href{mailto:bertrans@crm.umontreal.ca}{bertrans@crm.umontreal.ca}}

\Address{$^\ddag$~Department of Mathematics and Computer Science, Universit\'e du Qu\'ebec,\\
\hphantom{$^\ddag$}~Trois-Rivi\`eres, CP 500 (QC) G9A 5H7, Canada}
\EmailD{\href{mailto:grundlan@crm.umontreal.ca}{grundlan@crm.umontreal.ca}}

\Address{$^\S$~Centre de Recherches Math\'ematiques, Universit\'e de Montr\'eal,\\
\hphantom{$^\S$}~Montr\'eal CP 6128 (QC) H3C 3J7, Canada}
\EmailD{\href{mailto:hariton@crm.umontreal.ca}{hariton@crm.umontreal.ca}}

\ArticleDates{Received February 11, 2015, in f\/inal form June 09, 2015; Published online June 17, 2015}

\Abstract{The paper presents the bosonic and fermionic supersymmetric extensions of the structural equations describing conformally parametrized surfaces immersed in a Grasmann superspace, based on the authors' earlier results. A detailed analysis of the symmetry pro\-per\-ties of both the classical and supersymmetric versions of the Gauss--Weingarten equations is performed. A supersymmetric generalization of the conjecture establishing the necessary conditions for a system to be integrable in the sense of soliton theory is formulated and illustrated by the examples of supersymmetric versions of the sine-Gordon equation and the Gauss--Codazzi equations.}


\Classification{35Q51; 53A05; 22E70}

\renewcommand{\thefootnote}{\arabic{footnote}}
\setcounter{footnote}{0}

\section{Introduction}
Over the last decades, the concept of supersymmetry has been used extensively in particle physics and string theory \cite{BSTPV,Binetruy,Crombrugghe,Dine,Henkel06,Terning,Weinberg} as well as in hydrodynamic-type models \cite{Jackiw2,Das,Fatyga,GH11,Jackiw,MathieuLabelle,Manin,Mathieu}. Systems involving even and odd Grassmann variables are interesting because even Grassmann variables have properties similar to those of bosonic particles and odd Grassmann variables have properties similar to those of fermionic particles. These particles appear in the standard model, bosons as interaction particles and fermions as matter particles. Supersymmetric (SUSY) extensions have been constructed, for example, for the Korteweg--de Vries equation \cite{MathieuLabelle,Mathieu}, the Chaplygin gas equation in $(1+1)$- and $(2+1)$-dimensions (using parametrizations of the action for a superstring and a Nambu--Goto super\-membra\-ne, respectively) \cite{Jackiw}, the scalar Born--Infeld equation \cite{Hariton}, and the sine-Gordon equation \cite{Chaichian,Coleman,Grammaticos,GHS09,Siddiq05,Siddiq06,Witten}. Supersoliton solutions were obtained for a number of SUSY theories through a connection between the super-B\"acklund and super-Darboux transformations \cite{Aratyn,Chaichian,Gomes,GHS09,Liu,Siddiq06,Tian}. A Crum-type transformation was used to determine a number of supersoliton and multisupersoliton solutions, and the existence of inf\/initely many local conserved quantities was determined \cite{Grammaticos,MatveevSalle,Siddiq05}. In many cases, the integrability of supersymmetric systems has been demontrated by f\/inding Lax pairs and conservation laws \cite{MathieuLabelle,Mathieu}.

Superpositions of solutions of nonlinear SUSY systems are not as well understood as superpositions of solutions of nonlinear classical systems. As a result, SUSY dif\/ferential equations do not have as extensive a theoretical foundation as classical dif\/ferential equations. However, the method of prolongation of inf\/initesimal vector f\/ields for Lie symmetries, the methods for the classif\/ication of subalgebras and the symmetry reduction method can, to some extent, be adapted to the case of Grassmann-valued systems of dif\/ferential equations (see, e.g., \cite{AHW99,GHS09}).

A supersymmetric generalization of the structural equations (the Gauss, Codazzi and Ricci equations), constructed through the use of the exterior geometry formalism, was proposed in \cite{Bandos95,BSTPV,BSV95,Sorokin00}. This generalization was used to study superstrings and dif\/ferent super-$p$-branes. It was shown \cite{BSTPV}, using the superembedding approach, that these structural equations have their doubly supersymmetric counterparts.

\looseness=-1
The subject of our investigation is conformally parametrized surfaces immersed in a Grassmann superspace. This study is based on the methodology for the construction of SUSY extensions of the Gauss--Weingarten (GW) and Gauss--Codazzi (GC) equations developed in the authors' previous work \cite{BGH}. It involves the use of a moving frame formalism, leading to an explicit formulation of the structural equations for surfaces immersed in a Grassmann superspace. These equations constitute the SUSY extensions of the GW and GC equations. In \cite{BGH} we constructed two distinct extensions (one in terms of a bosonic superf\/ield and the other in terms of a fermionic superf\/ield) for each of these systems. For both SUSY extensions of the GC equations, Lie symmetry superalgebras were determined and the one-dimensional subalgebras of these superalgebras were classif\/ied into conjugacy classes under the action of their respective supergroups.

The main task undertaken in this paper is an analysis of the conditions for the existence of soliton and multisoliton solutions of the supersymmetric versions of dif\/ferential equations. For this purpose we adapt the symmetry group approach to the problem of integrability in the sense of soliton theory to the SUSY case. This approach proved to be ef\/fective in the classical case when it was f\/irst proposed in the form of a conjecture for point symmetries of the GW and GC equations by D.~Levi et al.~\cite{LST} and next developed by J.~Cie\'sli\'nski~\cite{Cieslinski92,CGS}. It establishes a~spectral technique which enables us to explicitly construct one-parameter families of surfaces associated with a given integrable system.

To formulate an analogue of the classical conjecture for the SUSY case we had to determine the symmetries of the GW equations for the classical case as well as for the bosonic and fermionic SUSY extensions and to compare them to the symmetries of the associated GC equations.
The conjecture states that, if the set of symmetries of the GC equations is larger than the set of symmetries of the GW equations, then we can introduce a spectral parameter into the GW equations and obtain a Lax pair associated with the GC equations, provided that the spectral parameter cannot be eliminated through a gauge transformation. This introduction can be done through the use of vector f\/ields that are symmetries of the original system, but not symmetries of the associated linear system.
We provide an algorithmic procedure for this analysis, facilitating the determination of the integrability of a system under consideration. We illustrate these results with the examples of the SUSY versions of the sine-Gordon equation and the GC equations.

The paper is organized as follows. In Section~\ref{section2}, we discuss the symmetry properties of the classical GW and GC equations, identify the Lie point symmetry algebras. Section~\ref{section3} is devoted to a brief outline of the properties of Grassmann variables and Grassmann algebras. In Section~\ref{section4}, we analyze bosonic and fermionic SUSY extensions of the GW and GC equations. In Section~\ref{section5}, we adapt the classical conjecture distinguishing integrable systems to the SUSY extensions of the GW and GC equations. Finally, in Section~\ref{section6}, we present possibilities for future research.

\section{Symmetries of the structural equations of conformally\\parametrized surfaces}\label{section2}
Consider a moving frame $\Omega$ on a smooth orientable conformally parametrized surface in 3-dimensional Euclidean space $\mathbb{R}^3$ which satisf\/ies the GW equations
\begin{gather}
\partial\left(\begin{matrix}
\partial F\\
\bar{\partial}F\\
N
\end{matrix}\right)=\left(\begin{matrix}
\partial u & 0 & Q \\
0 & 0 & \frac{1}{2}He^u \\
-H & -2e^{-u}Q & 0
\end{matrix}\right)\left(\begin{matrix}
\partial F\\
\bar{\partial}F\\
N
\end{matrix}\right), \qquad \partial\Omega=V_1\Omega,\nonumber\\
\bar{\partial}\left(\begin{matrix}
\partial F\\
\bar{\partial}F\\
N
\end{matrix}\right)=\left(\begin{matrix}
0 & 0 & \frac{1}{2}He^u \\
0 & \bar{\partial}u & \bar{Q} \\
-2e^{-u}\bar{Q} & -H & 0
\end{matrix}\right)\left(\begin{matrix}
\partial F\\
\bar{\partial}F\\
N
\end{matrix}\right), \qquad \bar{\partial}\Omega=V_2\Omega,
 \label{claGW}
\end{gather}
where we def\/ine the space $\mathcal{X}=(z,\bar{z})$ of independent variables, where $z=x+iy$ and $\bar{z}=x-iy$ are complex variables, and the space $\mathcal{U}=(H,Q,\bar{Q},u)$ of unknown functions. Here $\Omega=(\partial F,\bar{\partial}F,N)^{\rm T}$ is a moving frame of a conformally parametrized surface with the vector-valued function $F=(F_1,F_2,F_3)\colon \mathcal{R}\rightarrow\mathbb{R}^3$ (where $\mathcal{R}$ is a Riemann surface) satisfying the following normalization for the tangent vectors $\partial F$ and $\bar{\partial}F$ and the unit normal $N$
\begin{gather*}
\langle\partial F,\partial F\rangle=\langle\bar{\partial} F,\bar{\partial} F\rangle=0, \qquad\!\! \langle \partial F,\bar{\partial}F\rangle=\tfrac{1}{2}e^u,\qquad\!\!
\langle \partial F,N\rangle=\langle \bar{\partial} F,N\rangle=0, \qquad\!\! \langle N,N\rangle=1,\!
\end{gather*}
where the induced metric of the surface satisf\/ies $I=e^udzd\bar{z}$ with local $z$ and $\bar{z}$ coordinates on~$\mathcal{R}$. We have used the abbreviated notation
\begin{gather*}
\partial\equiv \partial_z=\tfrac{1}{2}(\partial_x-i\partial_y),\qquad \bar{\partial}\equiv\partial_{\bar{z}}=\tfrac{1}{2}(\partial_x+i\partial_y),
\end{gather*}
for the partial derivatives with respect to the complex variables~$z$ and~$\bar{z}$, respectively. The bracket $\langle\cdot,\cdot\rangle$ denotes the scalar product in 3-dimensional Euclidean space
\begin{gather}
\langle a,b\rangle=\sum_{i=1}^3a_ib_i,\qquad a,b\in\mathbb{C}^3.\label{inner}
\end{gather}
The quantities $Q$, $\bar{Q}$ and $H$ in equations \eqref{claGW} involve the second derivatives of the immersion function $F$ and are def\/ined as follows
\begin{gather*}
Q=\langle\partial^2F,N\rangle\in\mathbb{C},\qquad  H=2e^{-u}\langle\partial\bar{\partial}F,N\rangle\in\mathbb{R},
\end{gather*}
where the dif\/ferentials $Qdz^2$ and $\bar{Q}d\bar{z}^2$ def\/ined on the Riemann sphere~$\mathcal{R}$ are called Hopf dif\/ferentials while~$H$ is the mean curvature function of the surface.

The Gauss--Codazzi equations, which are the zero curvature condition (ZCC) for the potential matrices $V_1$ and $V_2$ taking values in a Lie algebra, are
\begin{gather*}
\bar{\partial}V_1-\partial V_2+[V_1,V_2]=0,
\end{gather*}
which reduce to the following three linearly independent equations
\begin{alignat}{3}
& \partial\bar{\partial} u+\tfrac{1}{2}H^2e^u-2Q\bar{Q} e^{-u}=0,\qquad && (\mbox{the Gauss equation}) & \nonumber\\
& \partial\bar{Q}=\tfrac{1}{2}e^{u}\bar{\partial}H,\qquad \bar{\partial}Q=\tfrac{1}{2}e^u\partial H. \qquad && (\mbox{the Codazzi equations})&
\label{claGC}
\end{alignat}
These equations guarantee the existence of conformally parametrized surfaces in~$\mathbb{R}^3$. A description of all inf\/initesimal symmetries of the GC equations was investigated~\cite{BGH} for conformally parametrized surfaces and the results can be summarized as follows.

In the case where the system of the GC equations has maximal rank over $M\subset \mathcal{X}\times \mathcal{U}$, it was found~\cite{BGH} that the set of all inf\/initesimal Lie point symmetries of the system forms an inf\/inite-dimensional Lie algebra~$\mathcal{L}_1$ spanned by the vector f\/ields
\begin{gather*}
X(\eta)=\eta(z)\partial_z+\eta'(z)(-2Q\partial_Q-U\partial_U),\qquad
Y(\zeta)=\zeta(\bar{z})\partial_{\bar{z}}+\zeta'(\bar{z})(-2\bar{Q}\partial_{\bar{Q}}-U\partial_U),\nonumber\\
e_0=-H\partial_H+Q\partial_Q+\bar{Q}\partial_{\bar{Q}}+2U\partial_U,
\end{gather*}
where $\eta$ and $\zeta$ are arbitrary functions of $z$ and $\bar{z}$ respectively, while $\eta'$ and $\zeta'$ are the derivatives of~$\eta$ and~$\zeta$ with respect to their arguments. Here and subsequently, we use the notation $U=e^u$. The generators $X(\eta)$ and $Y(\zeta)$ are two inf\/inite-dimensional families of conformal transformations, while $e_0$ is a dilation in the dependent variables which constitutes the center of the algebra.
The maximal f\/inite-dimensional subalgebra $L_1$ of the algebra $\mathcal{L}_1$ was obtained by expanding~$\eta$ and~$\zeta$ as power series with respect to their arguments. This algebra~$L_1$ is spanned by the seven generators
\begin{gather*}
e_0=-H\partial_H+Q\partial_Q+\bar{Q}\partial_{\bar{Q}}+2U\partial_U,\\
e_1=\partial_z,\qquad e_3=z\partial_z-2Q\partial_Q-U\partial_U,\qquad e_5=z^2\partial_z-4zQ\partial_Q-2zU\partial_U,\\
e_2=\partial_{\bar{z}},\qquad e_4=\bar{z}\partial_{\bar{z}}-2\bar{Q}\partial_{\bar{Q}}-U\partial_U,\qquad e_6=\bar{z}^2\partial_{\bar{z}}-4\bar{z}\bar{Q}\partial_{\bar{Q}}-2\bar{z}U\partial_U.
\end{gather*}

Let us now perform an analysis of the inf\/initesimal symmetries of the GW equations \eqref{claGW}. In the case where the system of GW equations has maximal rank over $M\subset \mathcal{X}\times \mathcal{U}$, the set of all inf\/initesimal symmetries of the system forms an inf\/inite-dimensional Lie algebra $\mathcal{L}_2$ spanned by the vector f\/ields
\begin{gather}
X(\eta)=\eta(z)\partial_z-\eta'(z)(U\partial_U+2Q\partial_Q),\qquad
Y(\zeta)=\zeta(\bar{z})\partial_{\bar{z}}-\zeta'(\bar{z})(U\partial_U+2\bar{Q}\partial_{\bar{Q}}),\nonumber\\
\hat{e}_0=-H\partial_H+Q\partial_Q+\bar{Q}\partial_{\bar{Q}}+2U\partial_U+F_i\partial_{F_i},\nonumber\\
T_i=\partial_{F_i},\qquad \mathcal{D}_i=F_i\partial_{F_{(i)}}+N_i\partial_{N_{(i)}},\qquad i=1,2,3,\nonumber\\
R_{ij}=(F_i\partial_{F_j}-F_j\partial_{F_i})+(N_i\partial_{N_j}-N_j\partial_{N_i}),\nonumber\\
S_{ij}=(F_i\partial_{F_j}+F_j\partial_{F_i})+(N_i\partial_{N_j}+N_j\partial_{N_i}),\qquad i< j=2,3.
\label{clavf}
\end{gather}
Here, we have used the notation $\eta'(z)=d\eta/dz$ and $\zeta'(\bar{z})=d\zeta/d\bar{z}$, where $\eta$ and $\zeta$ are arbitrary functions of $z$ and $\bar{z}$ respectively. The generators in \eqref{clavf} can be identif\/ied as follows: the $T_i$ generate translations in the $F_i$ directions respectively, $R_{ij}$ represent rotations in the direction of the $F_i$ and $N_i$ variables, $S_{ij}$ are local boost transformations and the vector f\/ields $e_0$, $\mathcal{D}_1$ and $\mathcal{D}_2$ correspond to scaling transformations. In addition, we obtain two inf\/inite-dimensional families of inf\/initesimal transformations generated by $X(\eta)$ and $Y(\zeta)$. The non-zero commutation relations between the generators \eqref{clavf} are
\begin{gather*}
[X(\eta_1),X(\eta_2)]=(\eta_1 \eta_2'-\eta_1'\eta_2)\partial_z+(\eta_1 ''\eta_2-\eta_1\eta_2 '')(U\partial_U+2Q\partial_Q),\\
{} [Y(\zeta_1),Y(\zeta_2)]=(\zeta_1\zeta_2'-\zeta_1'\zeta_2)\partial_{\bar{z}}+(\zeta_1 ''\zeta_2-\zeta_1\zeta_2 '')(U\partial_U+2\bar{Q}\partial_{\bar{Q}}),\\ {} [\hat{e}_0,T_i]=-T_i, \qquad [T_i,\mathcal{D}_j]=\delta_{ij}T_i, \qquad
[T_i,R_{jk}]=\delta_{ij}T_k-\delta_{ik}T_j,\\
{} [T_i,S_{jk}]=\delta_{ij}T_k+\delta_{ik}T_j,\qquad
[\mathcal{D}_i,R_{jk}]=\delta_{ij}S_{ik}-\delta_{ik}S_{ij},\qquad [\mathcal{D}_i,S_{jk}]=\delta_{ij}R_{ik}-\delta_{ik}R_{ji},\\ [R_{ij},S_{kl}]=\delta_{jk}S_{il}+\delta_{jl}S_{ik}-\delta_{ik}S_{jl}-\delta_{il}S_{jk},
\end{gather*}
where $\delta_{jk}$ is the Kronecker delta function. The Lie algebra $\mathcal{L}_2$ can be decomposed into the direct sum
\begin{gather*}
\mathcal{L}_2=\lbrace X(\eta)\rbrace\oplus\lbrace Y(\zeta)\rbrace\oplus\lbrace \hat{e}_0,T_i,\mathcal{D}_i,R_{ij},S_{ij}\rbrace,
\end{gather*}
which consists of two copies of the Virasoro algebra together with the 13-dimensional algebra generated by $\hat{e}_0$, $T_i$, $\mathcal{D}_i$, $R_{ij}$ and $S_{ij}$. If the functions $\eta$ and $\zeta$ are analytic, they can be expanded as power series with respect to~$z$ and~$\bar{z}$ respectively. The maximal f\/inite-dimensional subalgebra~$L_2$ of~$\mathcal{L}_2$ is spanned by the 19 generators
\begin{gather*}
\hat{e}_0=-H\partial_H+Q\partial_Q+\bar{Q}\partial_{\bar{Q}}+2U\partial_U+F_i\partial_{F_i},\\
e_1=\partial_z,\qquad e_3=z\partial_z-2Q\partial_Q-U\partial_U,\qquad e_5=z^2\partial_z-4zQ\partial_Q-2zU\partial_U,\\
e_2=\partial_{\bar{z}},\qquad e_4=\bar{z}\partial_{\bar{z}}-2\bar{Q}\partial_{\bar{Q}}-U\partial_U,\qquad e_6=\bar{z}^2\partial_{\bar{z}}-4\bar{z}\bar{Q}\partial_{\bar{Q}}-2\bar{z}U\partial_U,\\
T_i=\partial_{F_i},\qquad \mathcal{D}_i=F_i\partial_{F_{(i)}}+N_i\partial_{N_{(i)}},\qquad i=1,2,3,\\
R_{ij}=(F_i\partial_{F_j}-F_j\partial_{F_i})+(N_i\partial_{N_j}-N_j\partial_{N_i}),\\
S_{ij}=(F_i\partial_{F_j}+F_j\partial_{F_i})+(N_i\partial_{N_j}+N_j\partial_{N_i}),\qquad i< j=2,3,
\end{gather*}
which have the non-zero commutation relations
\begin{gather*}
[e_1,e_3]=e_1,\qquad [e_1,e_5]=2e_3,\qquad [e_3,e_5]=e_5,\\
{} [e_2,e_4]=e_2,\qquad [e_2,e_6]=2e_4,\qquad [e_4,e_6]=e_6,\\
{} [\hat{e}_0,T_i]=-T_i,\qquad [T_i,\mathcal{D}_j]=\delta_{ij}T_i,\qquad [T_i,R_{jk}]=\delta_{ij}T_k-\delta_{ik}T_j,\\
{} [T_i,S_{jk}]=\delta_{ij}T_k+\delta_{ik}T_j,\qquad  [\mathcal{D}_i,S_{jk}]=\delta_{ij}R_{ik}-\delta_{ik}R_{ji},\\
{} [\mathcal{D}_i,R_{jk}]=\delta_{ij}S_{ik}-\delta_{ik}S_{ij},\qquad [R_{ij},S_{kl}]=\delta_{jk}S_{il}+\delta_{jl}S_{ik}-\delta_{ik}S_{jl}-\delta_{il}S_{jk}.
\end{gather*}
The algebra $L_2$ can be decomposed as follows
\begin{gather*}
L_2=\lbrace e_1,e_3,e_5\rbrace\oplus\lbrace e_2,e_4,e_6\rbrace\oplus\lbrace T_i,\mathcal{D}_i,R_{ij},S_i\rbrace\ensuremath{\rlap{\raisebox{.15ex}{$\mskip
6.5mu\scriptstyle+ $}}\supset}\lbrace \hat{e}_0\rbrace.
\end{gather*}

In the theory of solitons, there exists a conjecture \cite{Cieslinski92,CGS,LST} to isolate integrable systems which states that this characterization can be performed by comparing the sets of symmetries of the original system and of its associated linear system. In the case where the sets of symmetries of both the original system and the non-parametric linear system (the GW system) are f\/inite-dimensional, we can compare the symmetries of the two systems by def\/ining the dif\/ferential projection operator $\pi$ as the following operator
\begin{gather*}
\pi(L_2)=L_2\omega,
\qquad \text{where}
\quad
\omega=z\partial+\bar{z}\bar{\partial}+H\partial_H+Q\partial_Q+\bar{Q}\partial_{\bar{Q}}+U\partial_U,
\end{gather*}
which involves all independent and dependent variables.
Here, $\omega$ is not necessarily an element of~$L_1$ or~$L_2$. The projection operator~$\pi$ has the property that~$\pi^n(L_2)=\pi(L_2)$ for any positive integer~$n$ and every element of the algebra~$L_2$. In fact, we have
\begin{gather*}
\pi^2(L_2)=\pi(L_2\omega)=L_2\omega^2=L_2\omega=\pi(L_2).
\end{gather*}

Under the above assumptions, the conjecture concerning integrable systems proposed in \cite{Cieslinski92,CGS,LST} can be formulated as follows.

\begin{conjecture}\label{conjecture2.1} \qquad
\begin{enumerate}\itemsep=0pt
\item[$1.$] In the case where $L_1=\pi(L_2)$, the original system is non-integrable in the sense of soliton theory. In the case where there exist reductions of the original system $($whose set of symmetries is $L_1')$ and the non-parametric linear system $($whose set of symmetries is $L_2')$ such that $L_1'\neq\pi(L_2')$, the reduced subsystem of the original system can be integrable.

\item[$2.$] In the case where $L_1\subset\pi(L_2)$, the system is a candidate to be integrable $($in the sense of soliton theory$)$ if it is possible to introduce a spectral parameter into the linear GW system, which represents a Lax pair, provided that the spectral parameter cannot be eliminated through a gauge transformation.
\end{enumerate}
\end{conjecture}

It should be noted that, under the above conjecture, the GC equations \eqref{claGC} do not form an integrable system since
$L_1=\pi(L_2)$.

\section{Certain aspects of Grassmann algebras}\label{section3}
We present a brief overview of the concepts related to Grassmann variables and Grassmann algebras. The formalism is based on the theory of supermanifolds as described, e.g., in~\cite{Berezin,BerezinMono,Binetruy,Cornwell,DeWitt,Freed,Kac,Rogers80,Rogers81,Varadarajan}. We consider a complex Grassmann algebra $\Gamma$ involving an arbitrary large (but f\/inite) number~$\mathfrak{k}$ of Grassmann generators ($\xi_1,\xi_2,\dots,\xi_{\mathfrak{k}}$). The exact number of generators is not essential as long as there is a suf\/f\/icient number of them to make all considered formulas meaningful. The Grassmann algebra $\Lambda$ can be decomposed into its even (bosonic) and odd (fermionic) parts
\begin{gather*}
\Lambda=\Lambda_{\rm even}+\Lambda_{\rm odd},
\end{gather*}
where $\Lambda_{\rm even}$ contains all terms involving a product of an even number of generators $\xi_k$, i.e., $1,\xi_1\xi_2,\xi_1\xi_3,\dots$, while $\Lambda_{\rm odd}$ contains all terms involving a product of an odd number of genera\-tors~$\xi_k$, i.e., $\xi_1,\xi_2,\xi_3,\dots,\xi_1\xi_2\xi_3,\dots$
The space~$\Lambda$ and/or~$\Lambda_{\rm even}$ replaces the f\/ield of complex numbers in the context of supersymmetry. The elements of~$\Lambda_{\rm even}$ and~$\Lambda_{\rm odd}$ are called even and odd supernumbers, respectively. An alternative decomposition for the Grassmann algebra~$\Lambda$ is
\begin{gather*}
\Lambda=\Lambda_{\rm body}+\Lambda_{\rm soul},
\end{gather*}
where
\begin{gather*}
\Lambda_{\rm body}=\Lambda^0[\xi_1,\xi_2,\dots,\xi_{\mathfrak{k}}]\simeq\mathbb{C},\qquad\Lambda_{\rm soul}=\sum_{k\geq1}\Lambda^k[\xi_1,\xi_2,\dots,\xi_{\mathfrak{k}}].
\end{gather*}
Here $\Lambda^0[\xi_1,\xi_2,\dots ,\xi_{\mathfrak{k}}]$ is used to refer to all elements of $\Lambda$ that do not involve any of the genera\-tors~$\xi_i$, while $\Lambda^k[\xi_1,\xi_2,\dots ,\xi_{\mathfrak{k}}]$ refers to all elements of $\Lambda$ that contain a product of $k$ generators (for instance, if we have $5$ generators $\xi_1,\xi_2,\xi_3,\xi_4,\xi_5$ then $\Lambda^2[\xi_1,\xi_2,\xi_3,\xi_4,\xi_5]$ refers to all terms involving $\xi_1\xi_2$, $\xi_1\xi_3$, $\xi_1\xi_4$, $\xi_1\xi_5$, $\xi_2\xi_3$, $\xi_2\xi_4$, $\xi_2\xi_5$, $\xi_3\xi_4$, $\xi_3\xi_5$ and $\xi_4\xi_5$). Because of the $\mathbb{Z}_0^+$-grading of the Grassmann algebra $\Lambda$, the bodiless elements in $\Lambda_{\rm soul}$ are non-invertible. Since the number~$\mathfrak{k}$ of Grassmann generators is f\/inite, it follows that the bodiless elements are nilpotent of degree at most~$\mathfrak{k}$.

In this paper, we use a $\mathbb{Z}_2$-graded complex vector space $V$ with even basis elements $u_i$, $i=1,2,\dots ,N$ and odd basis elements $v_\mu$, $\mu=1,2,\dots ,M$ and consider $W=\Lambda\otimes_{\mathbb{C}}V$. The even part of $W$
\begin{gather*}
W_{\rm even}=\left\lbrace\sum_ia_iu_i+\sum_\mu\underline{\alpha}_\mu v_\mu\,|\,  a_i\in\Lambda_{\rm even},\, \underline{\alpha}_\mu\in\Lambda_{\rm odd}\right\rbrace,
\end{gather*}
is a $\Lambda_{\rm even}$ module which can be identif\/ied with $\Lambda^{\times N}_{\rm even}\times\Lambda^{\times M}_{\rm odd}$ (which consists of $N$ copies of $\Lambda_{\rm even}$ and $M$ copies of $\Lambda_{\rm odd}$). To the original basis, consisting of the $u_i$ and $v_\mu$ (although $v_\mu\in\hspace{-0.35cm}\backslash\hspace{0.2cm} W_{\rm even}$), we associate the corresponding functionals
\begin{gather*}
 E_j\colon \ W_{\rm even}\rightarrow\Lambda_{\rm even}\colon \ E_j\left(\sum_ia_iu_i+\sum_\mu\underline{\alpha}_\mu v_\mu\right)=a_j,\\
 \Upsilon_\nu\colon \ W_{\rm even}\rightarrow\Lambda_{\rm odd}\colon \ \Upsilon_\nu\left(\sum_ia_iu_i+\sum_\mu\underline{\alpha}_\mu v_\mu\right)=\underline{\alpha}_\nu,
\end{gather*}
and view them as the coordinates (even and odd respectively) on $W_{\rm even}$. Any topological manifold locally dif\/feomorphic to a suitable $W_{\rm even}$ is called a supermanifold~\cite{Rogers80}.
Super-Minkowski space $\mathbb{R}^{(1,1\vert2)}$ is an example of such a supermanifold, being globally dif\/feomorphic to $\Lambda_{\rm even}^{\times2}\times\Lambda_{\rm odd}^{\times2}$, with bosonic light-cone coordinates $x_+$ and $x_-$, and fermionic coordinates $\theta^+$ and $\theta^-$. Therefore, $x_+$ and $x_-$ are linear combinations of terms containing an even number of generators: $1,\xi_1\xi_2,\xi_1\xi_3,\xi_1\xi_4,\dots ,\xi_2\xi_3,\xi_2\xi_4,\dots ,\xi_1\xi_2\xi_3\xi_4,\dots $ In contrast $\theta^+$ and $\theta^-$ are linear combinations of terms containing an odd number of generators : $\xi_1,\xi_2,\xi_3,\xi_4,\dots ,\xi_1\xi_2\xi_3,\xi_1\xi_2\xi_4,\xi_1\xi_3\xi_4,\xi_2\xi_3\xi_4,\dots$. Any fermionic (odd) variables~$\theta^+$ and $\theta^-$ satisfy the relation
\begin{gather}
(\theta^+)^2=(\theta^-)^2=\theta^+\theta^-+\theta^-\theta^+=0.\label{theta2}
\end{gather}
The supersymmetry transformations \eqref{susytrans} presented in the next section can be understood as changes in the coordinates of $\mathbb{R}^{(1,1\vert2)}$ which transform solutions of the SUSY GW equations and the SUSY GC equations, respectively, into solutions of the same equations in new coordinates for both the bosonic and fermionic SUSY extensions. A bosonic or fermionic smooth superf\/ield is a supersmooth~$G^\infty$ function from $\mathbb{R}^{(n_{\rm b}\vert n_{\rm f})}$ to $\Lambda$ (the values $n_{\rm b}$ and
$n_{\rm f}$ of the superspace $\mathbb{R}^{(n_{\rm b}\vert n_{\rm f})}$ stand for the number of bosonic and fermionic Grassmann coordinates respectively).

In this paper we use the convention that partial derivatives involving odd variables obey the following Leibniz rule for the product of two Grassmann-valued functions $h$ and $g$
\begin{gather*}
\partial_{\theta^\pm}(hg)=(\partial_{\theta^\pm}h)g+(-1)^{\deg(h)}h(\partial_{\theta^\pm}g),
\end{gather*}
where the degree of a homogeneous supernumber is given by
\begin{gather*}
\deg(h)=\begin{cases}
0& \mbox{if} \ h \ \mbox{is even},\\
1& \mbox{if} \ h \ \mbox{is odd}.
\end{cases}
\end{gather*}
We use the following ordering notation for partial derivatives $f_{\theta^+\theta^-}=\partial_{\theta^-}\partial_{\theta^+}f$. The partial derivatives with respect to  the fermionic coordinates satisfy $\partial_{\theta^i}\theta^j=\delta_i^j$, where $\delta_i^j$ is the Kronecker delta function and the indices~$i$ and $j$ each stand for $+$ or $-$. The fermionic operators~$\partial_{\theta^\pm}$,~$J_\pm$ and~$D_\pm$ in equations~\eqref{D} and~\eqref{J} alter the parity of a bosonic function to a fermionic function and vice versa. For instance, if $\phi$ is a bosonic function, then~$\partial_{\theta^+}\phi$ is an odd superf\/ield, while~$\partial_{\theta^+}\partial_{\theta^-}\phi$ is an even superf\/ield. For a Grassmann-valued composite function~$f(g(x_+))$, the chain rule is ordered as follows
\begin{gather*}
\frac{\partial f}{\partial x_+}=\frac{\partial g}{\partial x_+}\frac{\partial f}{\partial g}.
\end{gather*}
The interchange of mixed derivatives (with proper respect to the ordering of odd variables) is assumed throughout this paper. Additional details can be found in the books by Cornwell~\cite{Cornwell}, DeWitt~\cite{DeWitt}, Freed~\cite{Freed}, Kac~\cite{Kac}, Varadarajan~\cite{Varadarajan} and references therein.

\section{Supersymmetric versions of the Gauss--Weingarten\\ and Gauss--Codazzi equations}\label{section4}
In a previous paper \cite{BGH}, we constructed supersymmetric versions of the dif\/ferential equations which def\/ine surfaces in super-Minkowski space. These versions consisted of supersymmetric extensions of the Gauss--Weingarten and Gauss--Codazzi equations using bosonic and fermionic superf\/ields. The purpose of constructing such extensions was to construct surfaces immersed in a superspace ($\mathbb{R}^{(2,1\vert2)}$ for the bosonic extension and $\mathbb{R}^{(1,1\vert3)}$ for the fermionic extension).
We use the variables $x_\pm=\frac{1}{2}(t\pm x)$ which are the bosonic light-cone coordinates, and $\theta^\pm$ which are fermionic (anticommuting) variables satisfying \eqref{theta2}. Below, we present the outline of our procedure and its main results on which we base our further considerations.

Let $\mathcal{S}$ be a smooth orientable conformally parametrized surface immersed in the superspace given by a vector-valued superf\/ield $F(x_+,x_-,\theta^+,\theta^-)$ which, in view of \eqref{theta2}, can be decomposed~as
\begin{gather*}
F=F_m(x_+,x_-)+\theta^+\varphi_m(x_+,x_-)+\theta^-\psi_m(x_+,x_-)+\theta^+\theta^-G_m(x_+,x_-),\qquad m=1,2,3.
\end{gather*}
In the bosonic case, the functions $F_m$ and $G_m$ are bosonic-valued, while the functions~$\varphi_m$ and~$\psi_m$ are fermionic-valued. Conversely, in the fermionic case, the functions $F_m$ and $G_m$ are fermionic-valued, while the functions $\varphi_m$ and $\psi_m$ are bosonic-valued. In both cases, we def\/ine the covariant superspace derivatives to be
\begin{gather}
D_\pm=\partial_{\theta^\pm}-i\theta^\pm\partial_{x_\pm}.\label{D}
\end{gather}
The conformal parametrization of the surface $\mathcal{S}$ gives the following normalization on the superf\/ield
\begin{gather}
\langle D_iF,D_jF\rangle=fg_{ij},\qquad\langle D_iF,N\rangle=0,\qquad\langle N,N\rangle=1,\qquad i,j=1,2,\label{normalization}
\end{gather}
where $D_\pm F$ are the tangent vector superf\/ields and $N$ is a normal bosonic vector f\/ield which can be decomposed in the form
\begin{gather*}
N=N_m(x_+,x_-)+\theta^+\alpha_m(x_+,x_-)+\theta^-\beta_m(x_+,x_-)+\theta^+\theta^-H_m(x_+,x_-), \qquad m=1,2,3,
\end{gather*}
where $N_m$ and $H_m$ are bosonic functions, while $\alpha_m$ and $\beta_m$ are fermionic functions. In the bosonic case the function~$f$ which appears in \eqref{normalization} is a bodiless bosonic function (i.e., $f\in\Lambda_{\rm soul}$) of~$x_+$ and~$x_-$ which is a nilpotent function of some order~$k$. In the fermionic case the bosonic function $f$ may be bodiless or not. The values~1 and~2 of the indices $i$ and $j$ stand for~$+$ and~$-$ respectively. The bracket $\langle\cdot,\cdot\rangle$ denotes the scalar product~\eqref{inner} for 3-dimensional Euclidean space, where we use the property~\eqref{theta2} for any fermionic variables. This scalar product takes its values in the Grassmann algebra $\Lambda$. The coef\/f\/icients of the induced bosonic metric function~$g_{ij}$ on the surface~$\mathcal{S}$ are given by
\begin{gather*}
g_{ii}=0,\qquad g_{12}=\tfrac{1}{2}e^\phi,\qquad g_{21}=\tfrac{1}{2}\epsilon e^\phi,\qquad i=1,2,
\end{gather*}
where $\epsilon=1$ in the fermionic case and $\epsilon=-1$ in the bosonic case. It should be noted that the covariant metric tensor $g_{ij}$ is anti-symmetric in the indices $i$ and $j$ in the bosonic case while it is symmetric in those indices in the fermionic case. Here, the superf\/ield $\phi$ is assumed to be bosonic and can be expanded in terms of the fermionic variables $\theta^+$ and $\theta^-$:
\begin{gather*}
\phi=\phi_0(x_+,x_-)+\theta^+\phi_1(x_+,x_-)+\theta^-\phi_2(x_+,x_-)+\theta^+\theta^-\phi_3(x_+,x_-),
\end{gather*}
where $\phi_0$ and $\phi_3$ are bosonic functions while $\phi_1$ and $\phi_2$ are fermionic functions. The exponential function can be expanded as follows in terms of $\theta^+$ and $\theta^-$:
\begin{gather*}
e^{\pm\phi}=e^{\pm\phi_0}\big(1\pm\theta^+\phi_1\pm\theta^-\phi_2 \pm\theta^+\theta^-\phi_3-\theta^+\theta^-\phi_1\phi_2\big).
\end{gather*}
The SUSY extensions of the GW and GC equations are constructed in such a way that they are invariant under the transformations
\begin{gather}
x_\pm\rightarrow x_\pm+i\underline{\eta}_\pm\theta^\pm,\qquad \theta^\pm\rightarrow\theta^\pm+i\underline{\eta}_\pm,\label{susytrans}
\end{gather}
which are generated by the dif\/ferential SUSY operators
\begin{gather}
J_\pm=\partial_{\theta^\pm}+i\theta^\pm\partial_{x_\pm},\label{J}
\end{gather}
respectively. Here $\underline{\eta}_\pm$ are fermionic-valued parameters. The SUSY operators $J_\pm$ satisfy the following anticommutation relations:
\begin{gather*}
\lbrace J_n,J_m\rbrace=2i\delta_{mn}\partial_{x_m},\qquad\lbrace D_n,D_m\rbrace=-2i\delta_{mn}\partial_{x_m},\qquad\lbrace J_m,D_n\rbrace=0,\qquad m,n=1,2,\\
D_\pm^2=-i\partial_\pm,\qquad J_\pm^2=i\partial_\pm,
\end{gather*}
where $\delta_{mn}$ is the Kronecker delta function and the brace brackets denote anticommutation, unless otherwise specif\/ied. The values $1$ and $2$ of the indices $m$ and $n$ stand for $+$ and $-$ respectively. Here and subsequently, summation over repeated indices is understood.

In order to construct the SUSY version of the GW equations we assume that the second-order covariant derivatives of $F$ and the f\/irst-order covariant derivatives of the normal unit vector $N$ can be def\/ined in terms of the moving frame $\Omega=(D_+F,D_-F,N)^T$ on a surface $\mathcal{S}$, i.e.,
\begin{gather*}
D_jD_iF=\Gamma_{ij}^{\phantom{ij}k}D_kF+b_{ij}fN,\qquad
D_iN=b_i^{\phantom{i}k}D_kF+\omega_iN,\qquad i,j,k=1,2,
\end{gather*}
where the coef\/f\/icients $\Gamma_{ij}^{\phantom{ij}k}$ and~$\omega_i$ are fermionic
functions. The functions~$b_{ij}$ and~$b_i^{\phantom{i}k}$ are bosonic-valued in the bosonic extension and are fermionic-valued in the fermionic extension. Here, the values~$1$ and~$2$ of the indices~$i$, $j$ and $k$ stand for~$+$ and~$-$, respectively. We def\/ine the coef\/f\/icients~$b_{ij}$ to be
\begin{gather}
b_{11}=Q^+,\qquad b_{12}=-b_{21}=\tfrac{1}{2}e^\phi H,\qquad b_{22}=Q^-.\label{QH}
\end{gather}

In the bosonic extension, the moving frame $\Omega$ contains both bosonic and fermionic components. Under the above assumptions, we obtained the following results~\cite{BGH}

\begin{proposition}\label{proposition1}
For any bosonic vector superfields $F(x_+,x_-,\theta^+,\theta^-)$ and $N(x_+,x_-,\theta^+,\theta^-)$ satisfying the normalization conditions \eqref{normalization} and \eqref{QH}, the moving frame~$\Omega=(D_+F,D_-F,N)^{\rm T}$ on a smooth conformally parametrized surface immersed in the superspace $\mathbb{R}^{(2,1\vert2)}$ satisfies the SUSY GW equations
\begin{gather}
D_+\Omega=A_+\Omega, \qquad D_-\Omega=A_-\Omega,\nonumber \\
A_+=\left(\begin{matrix}
\Gamma_{11}^{\phantom{11}1} & \Gamma_{11}^{\phantom{11}2} & Q^+f \\
-\Gamma_{12}^{\phantom{12}1} & -\Gamma_{12}^{\phantom{12}2} & -\frac{1}{2}e^{\phi}Hf \\
H & 2e^{-\phi}Q^+ & 0
\end{matrix}\right), \qquad A_-=\left(\begin{matrix}
\Gamma_{12}^{\phantom{12}1} & \Gamma_{12}^{\phantom{12}2} & \frac{1}{2}e^\phi Hf \\
\Gamma_{22}^{\phantom{22}1} & \Gamma_{22}^{\phantom{22}2} & Q^-f \\
-2e^{-\phi}Q^- & H & 0
\end{matrix}\right).
\label{BGW}
\end{gather}
The zero curvature condition
\begin{gather}
D_+A_-+D_-A_+-\lbrace EA_+,EA_-\rbrace=0,\label{BZCC}
\end{gather}
where
\begin{gather*}
E=\pm\left(\begin{matrix}
1&0&0\\
0&1&0\\
0&0&-1
\end{matrix}\right),
\end{gather*}
constitutes the GC equations and corresponds to the following six linearly independent equations
\begin{alignat}{3}
& (i) \ && D_-\big(\Gamma_{11}^{\phantom{11}1}\big)+D_+\big(\Gamma_{22}^{\phantom{22}2}\big)+D_+\big(\Gamma_{12}^{\phantom{12}1}\big)
-D_-\big(\Gamma_{12}^{\phantom{12}2}\big)=0, &\nonumber\\
& (ii) && D_-\big(\Gamma_{11}^{\phantom{11}1}\big)-\Gamma_{11}^{\phantom{11}2}\Gamma_{22}^{\phantom{22}1} +D_+\big(\Gamma_{12}^{\phantom{12}1}\big)+\Gamma_{12}^{\phantom{12}2}\Gamma_{12}^{\phantom{12}1}+\frac{1}{2}H^2e^\phi f-2Q^+Q^-e^{-\phi}f=0, &\nonumber\\
& (iii) \ && Q^+\Gamma_{22}^{\phantom{22}2}-\Gamma_{11}^{\phantom{11}2}Q^-+D_-Q^+-Q^+D_-\phi+\frac{1}{2}e^\phi D_+H=0, &\nonumber\\
& (iv) && Q^-\Gamma_{11}^{\phantom{11}1}-\Gamma_{22}^{\phantom{22}1}Q^++D_+Q^--Q^-D_+\phi-\frac{1}{2}e^\phi D_-H=0, &\nonumber\\
& (v) && D_-\big(\Gamma_{11}^{\phantom{11}2}\big)-\Gamma_{12}^{\phantom{12}1}\Gamma_{11}^{\phantom{11}2} -\Gamma_{11}^{\phantom{11}2}\Gamma_{22}^{\phantom{22}2}-\Gamma_{11}^{\phantom{11}1}\Gamma_{12}^{\phantom{12}2} +D_+\big(\Gamma_{12}^{\phantom{12}2}\big)+2Q^+Hf=0, &\nonumber\\
& (vi) \ && D_+\big(\Gamma_{22}^{\phantom{22}1}\big)+\Gamma_{12}^{\phantom{12}2}\Gamma_{22}^{\phantom{22}1}-\Gamma_{22}^{\phantom{22}1} \Gamma_{11}^{\phantom{11}1}+\Gamma_{22}^{\phantom{22}2}\Gamma_{12}^{\phantom{12}1}-D_-\big(\Gamma_{12}^{\phantom{12}1}\big)+2Q^-Hf=0.&
 \label{BGC}
\end{alignat}
\end{proposition}

In the fermionic extension, the moving frame $\Omega$ contains only bosonic components. The fermionic counterpart of Proposition~\ref{proposition1} can be summarized as follows.

\begin{proposition}\label{proposition2}
For any fermionic vector superfield $F(x_+,x_-,\theta^+,\theta^-)$ and bosonic normal unit vector $N(x_+,x_-,\theta^+,\theta^-)$ satisfying the normalization conditions \eqref{normalization} and \eqref{QH}, the bosonic moving frame $\Omega=(D_+F,D_-F,N)^{\rm T}$ on a smooth conformally parametrized surface immersed in the superspace $\mathbb{R}^{(1,1\vert3)}$ satisfies the SUSY GW equations
\begin{gather}
D_+\left(\begin{matrix}D_+F\\D_-F\\N\end{matrix}\right)=\left(\begin{matrix}
\Gamma_{11}^{\phantom{11}1} & 0 & Q^+f \\
0 & 0 & -\frac{1}{2}e^\phi Hf \\
H & -2e^{-\phi}Q^+ & 0
\end{matrix}\right)\left(\begin{matrix}D_+F\\D_-F\\N\end{matrix}\right),\nonumber\\
D_-\left(\begin{matrix}D_+F\\D_-F\\N\end{matrix}\right)=\left(\begin{matrix}
0 & 0 & \frac{1}{2}e^\phi Hf \\
0 & \Gamma_{22}^{\phantom{22}2} & Q^-f \\
-2e^{-\phi}Q^- & -H & 0
\end{matrix}\right)\left(\begin{matrix}D_+F\\D_-F\\N\end{matrix}\right).
\label{FGW}
\end{gather}
The GC equations, which are equivalent to the ZCC
\begin{gather*}
D_+A_-+D_-A_+-\lbrace A_+,A_-\rbrace=0,
\end{gather*}
reduce to the following four linearly independent equations
\begin{alignat}{3}
&(i) && D_+\big(\Gamma_{22}^{\phantom{22}2}\big)+D_-\big(\Gamma_{11}^{\phantom{11}1}\big)=0,& \nonumber\\
&(ii) && D_-\big(\Gamma_{11}^{\phantom{11}1}\big)+2e^{-\phi}Q^+Q^-f=0,& \nonumber\\
&(iii) \ && D_+Q^--\frac{1}{2}e^\phi D_-H+Q^-\big(D_+\phi-\Gamma_{11}^{\phantom{11}1}\big)=0,& \nonumber\\
&(iv) && D_-Q^++\frac{1}{2}e^\phi D_+H+Q^+\big(D_-\phi-\Gamma_{22}^{\phantom{22}2}\big)=0.&\label{FGC}
\end{alignat}
\end{proposition}

\section{Conjecture on supersymmetric integrable systems}\label{section5}

In this section, we formulate a SUSY version of the Conjecture~\ref{conjecture2.1} on integrable systems described in Section~\ref{section2}. A symmetry supergroup $G$ of a SUSY system of equations consists of a local supergroup of transformations acting on a Cartesian product of supermanifolds $\mathcal{X}\times\mathcal{U}$, where~$\mathcal{X}$ is the space of four independent variables $(x_+,x_-,\theta^+,\theta^-)$ and $\mathcal{U}$ is the space of dependent superf\/ields.

Let $\mathcal{L}_1$ be a maximal f\/inite-dimensional superalgebra of Lie point symmetries associated with the system of nonlinear partial dif\/ferential equations (NPDEs) under consideration. Let~$\mathcal{L}_2$ be a~maximal f\/inite-dimensional superalgebra of Lie point symmetries of the linear system associated with the original system of NPDEs. Let~$\pi$ be a projection operator acting on the subalgebra~$\mathcal{L}_2$ such that
$
\pi(\mathcal{L}_2)=\mathcal{L}_2\omega$,
where $\omega$ is the dif\/ferential operator
\begin{gather*}
\omega=x_+\partial_{x_+}+x_-\partial_{x_-}+\theta^+\partial_{\theta^+}+\theta^-\partial_{\theta^-}+u^\alpha\partial_{u^\alpha} +\varphi^\beta\partial_{\varphi^\beta}
\end{gather*}
involving all independent bosonic and fermionic variables ($x_+,x_-,\theta^+,\theta^-$) and all dependent bosonic and fermionic superf\/ields, $u^\alpha$ and $\varphi^\beta$, respectively, appearing in the system of NPDEs. The common symmetries of the NPDEs and the linear spectral problem (LSP), associated with the original system of NPDEs, are the vector f\/ields which span the set
\begin{gather*}
\mathcal{L}_3=\mathcal{L}_1\cap\pi(\mathcal{L}_2)\neq\varnothing.
\end{gather*}
It should be noted that the set $\mathcal{L}_3$ is not necessarily an algebra.
The prolongation of one of these vector f\/ields acting on the LSP has to vanish for all wavefunctions of the LSP. In this case, the integrated form of a two-dimensional surface in a Lie algebra is given by the Fokas--Gel'fand immersion formula \cite{FG,FGFL,GPR}, whenever the tangent vectors on the surface are linearly independent.
Let us consider the set of vector f\/ields def\/ined by
\begin{gather*}
\mathcal{L}_4=\mathcal{L}_1\backslash\lbrace \mathcal{L}_1\cap\pi(\mathcal{L}_2)\rbrace.
\end{gather*}
Here, $\mathcal{L}_4$ consists of all symmetries of $\mathcal{L}_1$ that are not symmetries of $\mathcal{L}_2$. Again, $\mathcal{L}_4$ is not necessarily an algebra.
Under the above assumptions, an extension of the Conjecture~\ref{conjecture2.1} to SUSY integrable systems can be formulated as follows.

\begin{conjecture}\label{conjecture5.1}\quad
\begin{enumerate}\itemsep=0pt
\item[$1.$] If $\mathcal{L}_1=\pi(\mathcal{L}_2)$ then the system of NPDEs is not integrable.

\item[$2.$] If the following conditions are satisfied
\begin{enumerate}\itemsep=0pt
\item[$(a)$] $\pi(\mathcal{L}_2)$ is a proper subset of $\mathcal{L}_1$, that is
\begin{gather*}
\mathcal{L}_1\supset\pi(\mathcal{L}_2).
\end{gather*}
A free parameter can be introduced into the linear system using a symmetry transformation generated by one of the vector fields appearing in $\mathcal{L}_4$.

\item[$(b)$] The transformation given in $(a)$ acts in a nontivial way $($i.e., cannot be eliminated through an $\mathcal{L}_1$-valued gauge matrix function$)$.
\end{enumerate}
Then the system of NPDEs is a candidate to be an integrable system.
\end{enumerate}
\end{conjecture}

The proposed conjecture is illustrated through the following examples.

\begin{example}
The bosonic extension of the GC equations \eqref{BGC} involves eleven unknown functions $\mathcal{U}=(\phi,H,Q^+,Q^-,R^+,R^-,S^+,S^-,T^+,T^-,f)$, where $\phi$, $H$, $Q^+$, $Q^-$, $f$ are bosonic functions while $R^+$, $R^-$, $S^+$, $S^-$, $T^+$, $T^-$ are fermionic functions. In what follows we use the notation
\begin{gather*}
R^+=\Gamma_{11}^{\phantom{11}1},\qquad\! R^-=\Gamma_{11}^{\phantom{11}2},\qquad\! S^+=\Gamma_{12}^{\phantom{12}1},\qquad\! S^-=\Gamma_{12}^{\phantom{12}2},\qquad\! T^+=\Gamma_{22}^{\phantom{22}1},\qquad\! T^-=\Gamma_{22}^{\phantom{22}2}.
\end{gather*}
The action of the supergroup $G$ on the superf\/ields $\mathcal{U}$ of $(x_+,x_-,\theta^+,\theta^-)$ maps solutions of the bosonic version of the SUSY GC equations \eqref{BGC} to solutions of \eqref{BGC}. The bodiless bosonic function $f$ depends only on $x_+$ and $x_-$, in constrast with the other listed superf\/ields in $\mathcal{U}$ which can depend on $(x_+,x_-,\theta^+,\theta^-)$. Assuming that $G$ is a Lie supergroup as described in~\cite{Kac,Winternitz}, we found that its associated Lie superalgebra~$\mathfrak{g}_1$, whose elements are inf\/initesimal symmetries of the bosonic SUSY GC equations~\eqref{BGC}, was generated by the following eight vector f\/ields~\cite{BGH}
\begin{gather}
C_0= H\partial_H+Q^+\partial_{Q^+}+Q^-\partial_{Q^-}-2f\partial_f,\nonumber\\
K_0= -H\partial_H+Q^+\partial_{Q^+}+Q^-\partial_{Q^-}+2\partial_\phi,\nonumber\\
K_1^b=-2x_+\partial_{x_+}-\theta^+\partial_{\theta^+}+R^+\partial_{R^+}+2R^-\partial_{R^-}+S^-\partial_{S^-}-T^+\partial_{T^+}
+2Q^+\partial_{Q^+}+\partial_\phi,\nonumber\\
K_2^b=-2x_-\partial_{x_-}-\theta^-\partial_{\theta^-}-R^-\partial_{R^-}+S^+\partial_{S^+}+2T^+\partial_{T^+}+ T^-\partial_{T^-}+2Q^-\partial_{Q^-}+\partial_\phi,\nonumber\\
P_+=\partial_{x_+},\qquad  P_-=\partial_{x_-},\qquad
J_+=\partial_{\theta^+}+i\theta^+\partial_{x_+},\qquad  J_-=\partial_{\theta^-}+i\theta^-\partial_{x_-}.
\label{gen}
\end{gather}
The  generators $P_+$ and $P_-$ correspond to translations in the bosonic variables $x_+$ and $x_-$ respectively. We also have four dilations, of which two, $C_0$ and $K_0$, involve only the bosonic dependent variables, while the other two, $K_1^b$ and $K_2^b$, involve both independent and dependent, and both bosonic and fermionic variables. Finally, we also recover the two supersymmetry generators $J_+$ and $J_-$ which were identif\/ied previously in~\eqref{J}.

The algebra $\mathfrak{g}_1'$ of inf\/initesimal symmetries of the SUSY GW equations~\eqref{BGW} are spanned by the following vector f\/ields
\begin{gather*}
P_\pm=\partial_{x_\pm},\qquad J_\pm=\partial_{\theta^\pm}+i\theta^\pm\partial_{x_\pm},\qquad
\hat{C}_0=H\partial_H+Q^+\partial_{Q^+}+Q^-\partial_{Q^-}-2f\partial_f+N_i\partial_{N_i},\\
\hat{K}_0=-H\partial_H+Q^+\partial_{Q^+}+Q^-\partial_{Q^-}+2\partial_\phi-N_i\partial_{N_i},\\
K_1^b=-2x_+\partial_{x_+}-\theta^+\partial_{\theta^+}+R^+\partial_{R^+}+2R^-\partial_{R^-}+S^-\partial_{S^-}-T^+\partial_{T^+}+2Q^+\partial_{Q^+}+\partial_\phi,\\
K_2^b=-2x_-\partial_{x_-}-\theta^-\partial_{\theta^-}-R^-\partial_{R^-}+S^+\partial_{S^+}+2T^+\partial_{T^+}+T^-\partial_{T^-}+2Q^-\partial_{Q^-}+\partial_\phi,\\
G_i=F_i\partial_{F_i}+N_i\partial_{N_i},\qquad B_i=\partial_{F_i},\qquad\mbox{for} \quad i=1,2,3,\\
R_{ij}=F_i\partial_{F_j}-F_j\partial_{F_i}+N_i\partial_{N_j}-N_j\partial_{N_i},\qquad i<j=2,3.
\end{gather*}
Using the projection operator $\pi(\mathfrak{g}_1')=\mathfrak{g}_1'\omega$ involving all dependent and independent variables of the SUSY GC equations \eqref{BGC}, where
\begin{gather*}
\omega=x_+\partial_{x_+}+x_-\partial_{x_-}+\theta^+\partial_{\theta^+}+\theta^-\partial_{\theta^-}+\phi\partial_{\phi}+H\partial_H +Q^+\partial_{Q^+}+Q^-\partial_{Q^-}+f\partial_f\\
\hphantom{\omega=}{}+R^+\partial_{R^+}+R^-\partial_{R^-}+S^+\partial_{S^+}+S^-\partial_{S^-}+T^+\partial_{T^+} +T^-\partial_{T^-},
\end{gather*}
and  comparing the resulting vector f\/ields with the generators of $\mathfrak{g}_1$, given by~\eqref{gen}, we conclude that $\mathfrak{g}_1=\pi(\mathfrak{g}_1')$, which implies that the SUSY GC equations are non-integrable as in the classical case.
\end{example}

\begin{example}
As another example, we apply the conjecture to the SUSY sine-Gordon equation as formulated in~\cite{Chaichian}
\begin{gather}
D_+D_-\Phi=i\sin\Phi,\label{sG}
\end{gather}
where $\Phi$ is a bosonic superf\/ield. Its Lie symmetry superalgebra $\mathfrak{g}_3$ is spanned by the vector f\/ields
\begin{gather}
P_\pm=\partial_{x_\pm},\qquad J_\pm=\partial_{\theta^\pm}+i\theta^\pm\partial_{x_\pm},\qquad \mathcal{K}=2x_+\partial_{x_+}-2x_-\partial_{x_-}+\theta^+\partial_{\theta^+}-\theta^-\partial_{\theta^-}.\label{sGsym}
\end{gather}
The non-parametric linear problem (the GW equations) associated with the SUSY sine-Gordon equation \eqref{sG} is given by
\begin{gather}
D_\pm\Psi=B_\pm\Psi,\qquad\mbox{where}  \quad \Psi=\left(\begin{matrix}
\psi_{11} & \psi_{12} & f_{13} \\
\psi_{21} & \psi_{22} & f_{23} \\
f_{31} & f_{32} & \psi_{33}
\end{matrix}\right),\nonumber\\
B_+=\frac{1}{2}\left(\begin{matrix}
0 & 0 & ie^{i\Phi} \\
0 & 0 & -ie^{-i\Phi} \\
-e^{-i\Phi} & e^{i\Phi} & 0
\end{matrix}\right), \qquad B_-=\left(\begin{matrix}
iD_-\Phi & 0 & -i \\
0 & -iD_-\Phi & i \\
-1 & 1 & 0
\end{matrix}\right).
\label{sGLP}
\end{gather}
Here the $\psi_{ij}$ are bosonic superf\/ields and the~$f_{ij}$ are fermionic superf\/ields, $i,j=1,2,3$. The inf\/initesimal symmetry generators $\mathfrak{g}_3'$ of equations~\eqref{sGLP} are spanned by the vector f\/ields
\begin{gather*}
P_\pm=\partial_{x_\pm},\qquad J_\pm=\partial_{\theta^\pm}+i\theta^\pm\partial_{x_\pm}, \qquad  G_1=\psi_{11}\partial_{\psi_{11}}+\psi_{21}\partial_{\psi_{21}}+f_{31}\partial_{f_{31}},\\
G_2=\psi_{12}\partial_{\psi_{12}}+\psi_{22}\partial_{\psi_{22}}+f_{32}\partial_{f_{32}}, \qquad G_3=f_{13}\partial_{f_{13}}+f_{23}\partial_{f_{23}}+\psi_{33}\partial_{\psi_{33}}.
\end{gather*}
Using the projection operator $\pi$ def\/ined as $\pi(\mathfrak{g}_3')=\mathfrak{g}_3'\omega$, where
\begin{gather*}
\omega=x_+\partial_{x_+}+x_-\partial_{x_-}+\theta^+\partial_{\theta^+}+\theta^-\partial_{\theta^-}+\Phi\partial_\Phi,
\end{gather*}
we obtain the relation $\mathfrak{g}_3\supset\pi(\mathfrak{g}_3')$, which implies that the SUSY sine-Gordon equation may be integrable, as in the classical case. The fact that the generator $\mathcal{K}$ in~\eqref{sGsym} does not appear in the symmetries of the linear problem~\eqref{sGLP} of the SUSY sine-Gordon equations \eqref{sG} allows us to introduce a bosonic spectral parameter $\lambda$ through $\mathcal{K}$. This is accomplished by introducing a one-parameter group associated with the dilation $\mathcal{K}$ through the transformation $\tilde{x}_+=\lambda x_+$, $\tilde{x}_-=\lambda^{-1}x_-$, $\tilde{\theta}^+=\lambda^{1/2}\theta^+$ and $\tilde{\theta}^-=\lambda^{-1/2}\theta^-$, $\lambda=\pm e^{\mu}$, where $\mu\in\Lambda_{\rm even}$, into the linear system~\eqref{sGLP} which gives us
\begin{gather}
D_+\Psi=B_+\Psi, \qquad D_-\Psi=B_-\Psi,\nonumber\\
B_+=\frac{1}{2\sqrt{\lambda}}\left(\begin{matrix}
0 & 0 & ie^{i\Phi} \\
0 & 0 & -ie^{-i\Phi} \\
-e^{-i\Phi} & e^{i\Phi} & 0
\end{matrix}\right), \qquad B_-=\left(\begin{matrix}
iD_-\Phi & 0 & -i\sqrt{\lambda} \\
0 & -iD_-\Phi & i\sqrt{\lambda} \\
-\sqrt{\lambda} & \sqrt{\lambda} & 0
\end{matrix}\right),
 \label{sGLSP}
\end{gather}
which coincide with the results found in~\cite{Siddiq05}. The ZCC of equation \eqref{sGLSP} takes the form~\eqref{BZCC}, where the matrices~$A_+$ and~$A_-$ are replaced by the matrices~$B_+$ and~$B_-$, respectively. The connection between the super-Darboux transformations and the super-B\"acklund transformations for the sine-Gordon equation~\eqref{sG} allows the construction of explicit multi-super-soliton solu\-tions~\cite{Chaichian,Grammaticos,Siddiq06}.
\end{example}

\begin{example}
The fermionic extension of the GC equations \eqref{FGC} involves seven unknown functions $\mathcal{U}=(\phi,H,Q^+,Q^-,R^+,$ $T^-,f)$ where $\phi$ and $f$ are bosonic functions, while $H$, $Q^+$, $Q^-$, $R^+=\Gamma_{11}^{\phantom{11}1}$, $T^-=\Gamma_{22}^{\phantom{22}2}$ are fermionic functions. Proceeding in a similar manner as in the bosonic SUSY case, we obtain a Lie symmetry superalgebra $\mathfrak{g}_2$ consisting of the following six bosonic inf\/initesimal generators \cite{BGH}
\begin{gather*}
P_+=\partial_{x_+},\qquad P_-=\partial_{x_-},\\
C_0=H\partial_H+Q^+\partial_{Q^+}+Q^-\partial_{Q^-}-2f\partial_f,\\
K_0=-H\partial_H+Q^+\partial_{Q^+}+Q^-\partial_{Q^-}+2\partial_\phi,\\
K_1^f=-2x_+\partial_{x_+}-\theta^+\partial_{\theta^+}+2Q^+\partial_{Q^+}+R^+\partial_{R^+}+\partial_\phi,\\
K_2^f=-2x_-\partial_{x_-}-\theta^-\partial_{\theta^-}+2Q^-\partial_{Q^-}+T^-\partial_{T^-}+\partial_\phi,
\end{gather*}
together with the three fermionic generators
\begin{gather*}
J_+=\partial_{\theta^+}+i\theta^+\partial_{x_+},\qquad J_-=\partial_{\theta^-}+i\theta^-\partial_{x_-},\qquad W=\partial_H.
\end{gather*}
The generators $W$ and $P_\pm$ correspond to translations in the fermionic variable $H$ and the bosonic variables~$x_\pm$ respectively. We obtain two dilations~$C_0$ and~$K_0$ involving only dependent variables, together with two additional dilations~$K_1^f$ and~$K_2^f$, which involve both dependent and independent variables. We also recover the two supersymmetry generators~$J_+$ and~$J_-$.

The Lie symmetry algebra $\mathfrak{g}_2'$ of the SUSY GW equations \eqref{FGW} is spanned by the vector f\/ields
\begin{gather*}
P_\pm=\partial_{x_\pm},\qquad J_\pm=\partial_{\theta^\pm}+i\theta^\pm\partial_{x_\pm},\\
\hat{C}_0=H\partial_H+Q^+\partial_{Q^+}+Q^-\partial_{Q^-}-2f\partial_f+N_i\partial_{N_i},\\
\hat{K}_0=-H\partial_H+Q^+\partial_{Q^+}+Q^-\partial_{Q^-}+2\partial_\phi-N_i\partial_{N_i},\\
K_1^f=-2x_+\partial_{x_+}-\theta^+\partial_{\theta^+}+2Q^+\partial_{Q^+}+R^+\partial_{R^+}+\partial_\phi,\\
K_2^f=-2x_-\partial_{x_-}-\theta^-\partial_{\theta^-}+2Q^-\partial_{Q^-}+T^-\partial_{T^-}+\partial_\phi,\\
G_i=F_i\partial_{F_i}+N_i\partial_{N_i},\qquad B_i=\partial_{F_i},\qquad\mbox{for} \quad i,j=1,2,3,\\
R_{ij}=F_i\partial_{F_j}-F_j\partial_{F_i}+N_i\partial_{N_j}-N_j\partial_{N_i},\qquad i<j=2,3.
\end{gather*}
By using the projector $\pi$, def\/ined as $\pi(\mathfrak{g}_2')=\mathfrak{g}_2'\omega$, where
\begin{gather*}
\begin{split}
&\omega=x_+\partial_{x_+}+x_-\partial_{x_-}+\theta^+\partial_{\theta^+}+\theta^-\partial_{\theta^-}+H\partial_H +Q^+\partial_{Q^+} +Q^-\partial_{Q^-}\\
&\hphantom{\omega=}{} +\phi\partial_\phi+f\partial_f+R^+\partial_{R^+}+T^-\partial_{T^-},
\end{split}
\end{gather*}
we obtain that the set of symmetries of the SUSY GW equations \eqref{FGW} is a proper subset of the set of Lie symmetries of the SUSY GC equations~\eqref{FGC}. More specif\/ically the translation in~$H$ generated by~$W$ is not a symmetry of the SUSY GW equations. Therefore we can introduce a~fermionic parameter $\underline{\lambda}$ in the SUSY GW equations~\eqref{FGW} with the potential matrices
\begin{gather*}
A_+=\left(\begin{matrix}
R^+ & 0 & Q^+f \\
0 & 0 & -\frac{1}{2}e^\phi(H+\underline{\lambda})f \\
H+\underline{\lambda} & -2e^{-\phi}Q^+ & 0
\end{matrix}\right),\\
A_-=\left(\begin{matrix}
0 & 0 & \frac{1}{2}e^\phi(H+\underline{\lambda})f \\
0 & T^- & Q^-f \\
-2e^{-\phi}Q^- & -(H+\underline{\lambda}) & 0
\end{matrix}\right).
\end{gather*}
The parameter $\underline{\lambda}$ cannot be eliminated through a gauge transformation. This suggests that the fermionic version of the SUSY GC equations~\eqref{FGC} may be integrable.
\end{example}

\section{Concluding remarks and outlook}\label{section6}

\looseness=1
The objective of this paper was to compare the symmetries of the SUSY GW equations with those of the SUSY GC equations for both the bosonic and fermionic extensions. This comparison allowed us to formulate a generalization of the conjecture establishing the necessary conditions for a system to be integrable in the sense of soliton theory. The symmetry analysis developed in this paper could be extended in several directions. First, it should be noted that the list of symmetries for the bosonic and fermionic SUSY structural equations found in this paper is not necessarily exhaustive since the symmetry criterion has not been proven for equations involving Grassmann variables. A~comprehensive list of all symmetries of the bosonic and fermionic SUSY Gauss--Weingarten and SUSY Gauss--Codazzi equations could be compiled. This would require the development of a computer Lie algebra symmetry packa\-ge capable of handling both bosonic and fermionic symmetries. Another possibility would be to extend the procedure to hypersurfaces in higher dimensions. It could also be worth attempting to establish a SUSY version of Noether's theorem in order to determine conserved quantities, and to derive a SUSY version of the Weierstrass--Enneper formula for the immersion of surfaces in a multidimensional superspace. One could also investigate how characteristics associated with integrable models such as Hamiltonian structures and conserved quantities manifest themselves in the SUSY case. Another worthwhile subject is a variational problem of geometric functionals (e.g., Willmore functionals), which can be interpreted as actions from which we can determine the Euler--Lagrange equations for a given surface immersed in a superspace. Recurrence operators of generalized symmetries of the SUSY GC equations could be used to obtain the recurrence relations for the surfaces. A~complete invariant geometrical characterization of these surfaces in the superspace remains to be done. A~singularity analysis of the SUSY system under consideration could be performed in connection with Lie groups in order to verify the Painlev\'e property. This would be motivated by the goal of obtaining explicit analytic solutions. Such analytic solutions can be useful for observing the qualitative behaviour of solutions which would otherwise be dif\/f\/icult to detect numerically. The existence of dif\/ferent types of soliton solutions constitutes such an example. An essential step in the further development of the theory of surfaces associated with SUSY integrable systems would be a generalization of the known formulas for constructing soliton surfaces immersed in Lie algebras, namely the Sym--Tafel \cite{Sym,Tafel} and the Fokas--Gel'fand \cite{FGFL} formulas. Our procedure for introducing a spectral parameter in the GW equations could make this task feasible.

\subsection*{Acknowledgements}

We thank professor D.~Levi (University of Roma Tre) for useful discussions on this topic. AMG's work was supported by a research grant from NSERC. SB~acknowledges a doctoral fellowship provided by the FQRNT of the Gouvernement du Qu\'ebec. AJH~wishes to acknowledge and thank the Mathematical Physics Laboratory of the Centre de Recherches Math\'ematiques for the opportunity to contribute to this research.

\pdfbookmark[1]{References}{ref}
\LastPageEnding

\end{document}